\documentclass{article}
\usepackage{floatflt}
\usepackage[hang,small,bf]{caption}
\usepackage{amssymb}
\title{LHCb VELO software alignment\\
       {\large Part III: the alignment of the relative sensor positions}}
\author{M. Gersabeck{\address[GLA]{University of Glasgow}},
C. Parkes{\addressmark[GLA]},
S. Viret{\addressmark[GLA]}
}

\usepackage{glas}

\begin{document}

\begin{titlepage}{GLAS-PPE/2008-05}{15$^{\underline{\rm{th}}}$ April 2008}
\title{LHCb VELO software alignment, Part III: the alignment of the relative sensor positions}

\author{M. Gersabeck$^1$, C. Parkes$^1$, S. Viret$^1$\\
\\
$^1$ University of Glasgow, Glasgow, G12 8QQ, Scotland}

\begin{abstract}
The LHCb Vertex Locator contains 42 silicon sensor modules. Each module has two silicon sensors. A method for determining the relative alignment of the silicon sensors within each module from data is presented. The software implementation details are discussed. Monte-Carlo simulation studies are described that demonstrate an alignment precision of $1.3~\mathrm{\mu{}m}$ is obtained in the sensor plane.
\end{abstract}

\vspace{0.5cm}
\begin{center}
{\em LHCb Public Note, LHCb-2007-138}
\end{center}

\newpage
\end{titlepage}

\maketitle

\begin{abstract}
The LHCb Vertex Locator contains 42 silicon sensor modules. Each module has two silicon sensors. A method for determining the relative alignment of the silicon sensors within each module from data is presented. The software implementation details are discussed. Monte-Carlo simulation studies are described that demonstrate an alignment precision of $1.3~\mathrm{\mu{}m}$ is obtained in the sensor plane.
\end{abstract}


\tableofcontents

\listoffigures

\section{Introduction}
High precision spatial resolution of the vertex detector (VELO, {\bf \cite{bib:ReOptTDR-03}}) is essential for the success of LHCb's B-physics programme.
To achieve this, the alignment of each active part of the detector should be determined with an accuracy significantly better than the single hit resolution.

The algorithms required to align the VELO modules relative to each other within each VELO half {\bf \cite{bib:Vir-05}}, and to align the two VELO halves with respect to each other {\bf \cite{bib:Vir-07}} have been described in the Glasgow alignment group's notes.

In these notes it was assumed that the knowledge of the relative position of the $R$ and $\Phi$ sensors on each module would rely only on metrology information. 
This would imply that during operation the relative sensor position cannot be monitored or corrected for. 
Furthermore, as the metrology precision is of the order of the best single hit precision of a sensor\footnote{about $5~\mathrm{\mu{}m}$ for the smallest pitch and optimal track angle} it would slightly degrade the tracking and vertexing performance. 

In this note, a novel algorithm to obtain the relative sensor misalignment from data is presented. Its implementation is described in section~\ref{sec:alignment} and simulation test results are discussed in section~\ref{sec:MC} before concluding in section~\ref{sec:concl}.

\section{The VELO alignment procedure}
\label{sec:alignment}

The alignment of the LHCb VELO will proceed in a number of stages.
The sensors were built and measured~\cite{bib:metr-07,bib:metr-07b} with high precision.
They were subsequently tested and then assembled onto their supporting structures.
Further metrology was performed on the two detector halves~\cite{bib:metr-08} before insertion into the LHCb experiment.
During data-taking, a track-based software alignment will be performed after each re-insertion of the two VELO halves.
Relevant observables will be monitored continuously to ensure a high performance at all times.
Finally, an annual re-processing is foreseen which gives the opportunity to further refine the alignment constants.

\subsection{The software alignment strategy}
\label{sec:align}

The VELO software alignment is divided into three parts, all based on track residuals. 
The two stages described in previous notes, allow the alignment of the VELO modules relative to each other within the two VELO halves, followed by the relative alignment of the two halves. 
The algorithm described here to extract the relative misalignment of the $R$ and $\Phi$ sensors on each module will precede the other two steps of the alignment procedure.

Since the two silicon sensors are glued on a single substrate~\cite{bib:glue-07} it is not expected that the sensors will move significantly with respect to each other. 
Hence, this procedure to determine the relative sensor alignment must be performed on start-up to retrieve the initial position and thereafter at regular intervals to monitor the position, but it does not need not need to be performed each fill.

\subsubsection{The VELO half and module alignment}
Both the VELO half and the module alignment use the same global minimization technique.
Here, \emph{global} means that both the alignment parameters and track parameters are fitted in one single step.
Such a procedure involves the inversion of a very large matrix.
This is done via the Millepede {\bf \cite{bib:Blo-07}} algorithm, which has already been successfully used in other HEP experiments. 
The global inversion method has the advantages of being both fast and of providing an easy way to constrain the system against global linear deformations while taking into account correlations between the various detector elements.

\vspace{-6pt}
\subsubsection{A method to determine sensor-to-sensor misalignments}
\vspace{-2pt}
For the determination of the relative misalignment of $R$ and $\Phi$ sensors the problem is no longer linearisable, which is essential for exploiting a global matrix inversion technique.
Hence, an iterative approach is used here that extracts the misalignment constants from the distribution of residuals plotted against position.

\begin{figure}
\centering
\includegraphics[width=0.45\textwidth]{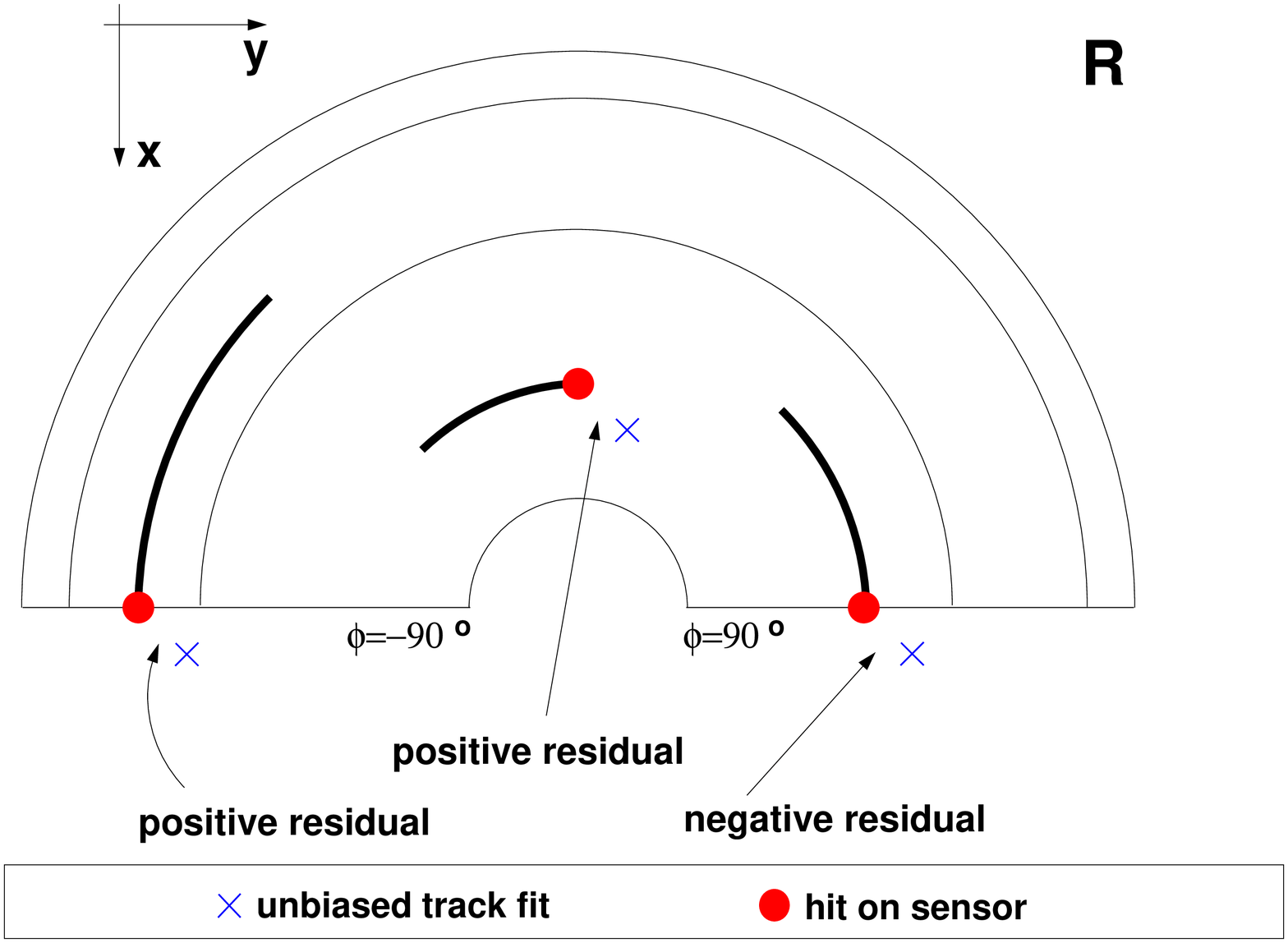}
\includegraphics[width=0.45\textwidth]{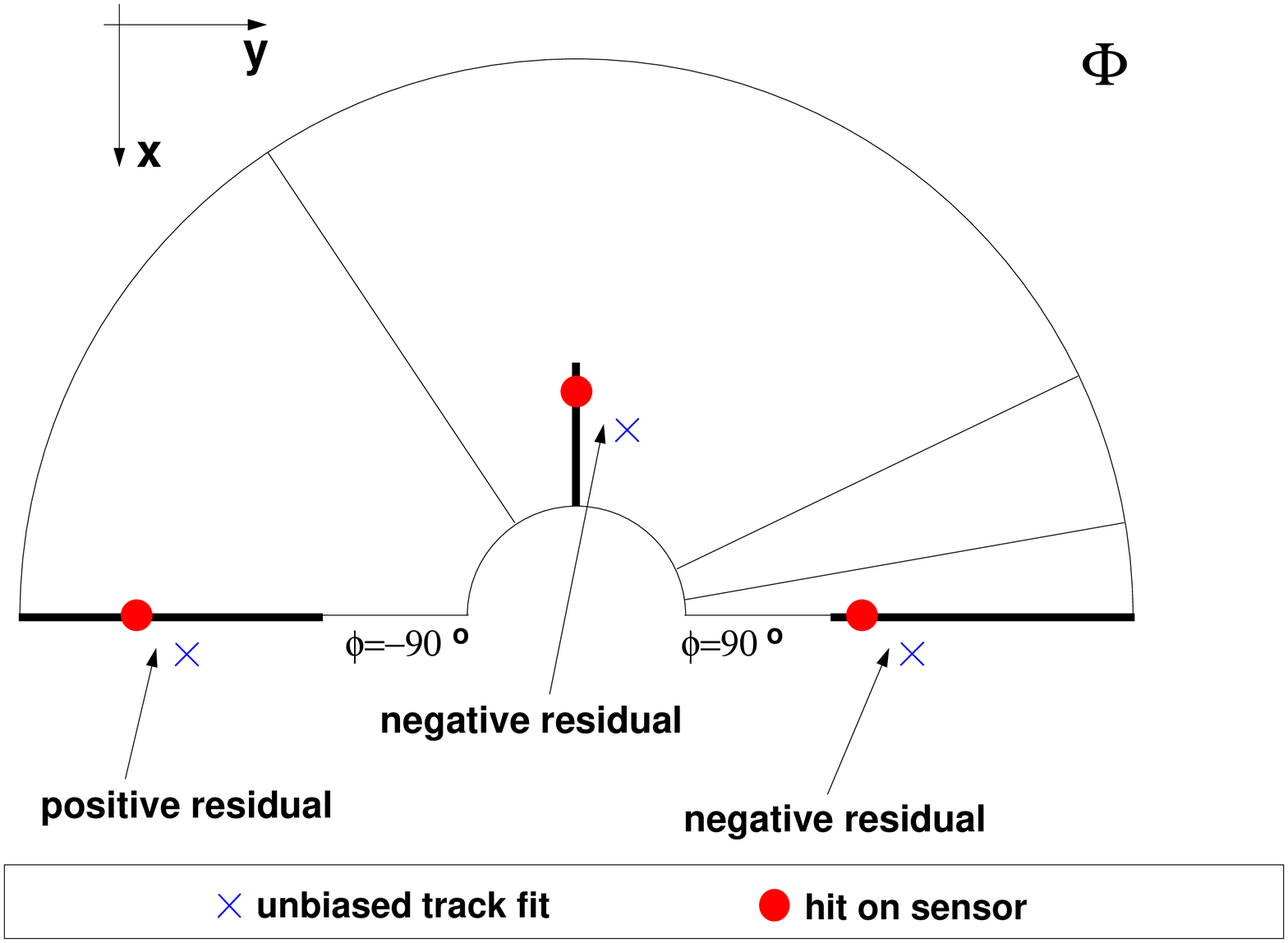}
\caption{Influence of misalignments on residuals of $R$ and $\Phi$ sensors. The misalignment shown is a translation both along the negative $x$ and $y$ direction.}
\label{fig:residuals}
\end{figure}

The characteristic shape of these distributions can easily be related to the misalignment of the sensors.
In a non-misaligned geometry, the plane of the sensor surface is, to first order approximation, the $x$-$y$ coordinate plane in the LHCb coordinate system.
The $y$-axis runs along the straight edge of the sensor, while the $x$-axis lies on the symmetry axis of the sensor and defines $\phi=0$ (see fig.~\ref{fig:residuals}).
To further reduce the risk of contact with the RF-foil the sensors are slightly tilted inwards towards their straight edge by about $2~\mathrm{mrad}$.

Clearly, measurements will only be affected by misalignment translations that are non-parallel to the corresponding strip on the sensor.
Thus, $R$ sensors are most sensitive to $x$-translations around $\phi=0$, whereas they are most sensitive to $y$-translations near $\phi=\pm\pi/2$.
The opposite is true for $\Phi$ sensors (see fig.~\ref{fig:residuals}).

Ideally, the method described below should be applied for each sensor in its local co-ordinate system, as it is sensitive to translations of the sensor in its own plane.
However, to simplify the fit code, all fits are done in a common co-ordinate system.
For this the respective VELO half co-ordinate system has been chosen as it also allows the algorithm to work when the VELO halves are retracted.
The simplification of using a common co-ordinate system is justified as explained in the following.

Defining the residual as the difference between the hit position\footnote{Here, the residual is calculated (as provided by the {\tt DeVelo[R/Phi]Type} classes) as the perpendicular distance to the strip hit in the sensor plane including inter-strip fractions (as provided by the {\tt VeloClusterPositionTool} \cite{bib:Szu-07}).} and the extrapolated position of an unbiased track fit one can write the relation between misalignments ($\Delta_{i}$) and residuals ($\epsilon_{R/\Phi}$) as follows.
Note that the track position is only extrapolated to the $z$-position of the sensor, i.e. neglecting the sensor tilts around the $x$ and $y$ axes.
However, this effect is only of the order of the square of the tilts, hence in the sub-micron range.

\begin{equation}
\begin{array}{lcl}
\epsilon_{R} = -\Delta_{x}\cos\phi_{track}+\Delta_{y}\sin\phi_{track} & &(R\mathrm{~sensor}),\\
\epsilon_{\Phi} = +\Delta_{x}\sin\phi_{cluster}+\Delta_{y}\cos\phi_{cluster}+\Delta\gamma r_{track} & & (\Phi\mathrm{~sensor}),
\label{eqn:residual}
\end{array}
\end{equation}
where $\Delta_{\gamma}$ describes a misalignment in the form of a rotation around the $z$ axis, which translates into a shift in $\phi$ by multiplication with the radial co-ordinate of the extrapolated track in the sensor plane.
As the $\Delta_{\gamma}$ term does not contain any $\phi$ dependence it is sufficient to leave it as a free parameter in the form of a constant when fitting the shape of the residual distribution as a function of $\phi$.

The sensor tilts around the $x$ and $y$ axes are neglected again as the residuals are determined in the sensor plane but plotted against $r$ and $\phi$ in the respective VELO half co-ordinate system.
Once more, this is justified as their effect on $x$- and $y$-translations is only of second order. 

The value for $\Delta_\gamma$ can be directly extracted by fitting the residual distribution on the $\Phi$~sensor versus $r$ (rather than versus $\phi$ as just discussed). Fitting a linear function to the residual distribution  versus $r$ gives $\Delta_\gamma$ as the slope. This value is used in the iterations of this sensor alignment procedure to improve the convergence of the algorithm. The final value for the $z$-rotation alignment constant will not be determined by this method, but is determined by the module alignment algorithm as reported in our previous notes.

\begin{figure}
\centering
\includegraphics[width=0.9\textwidth]{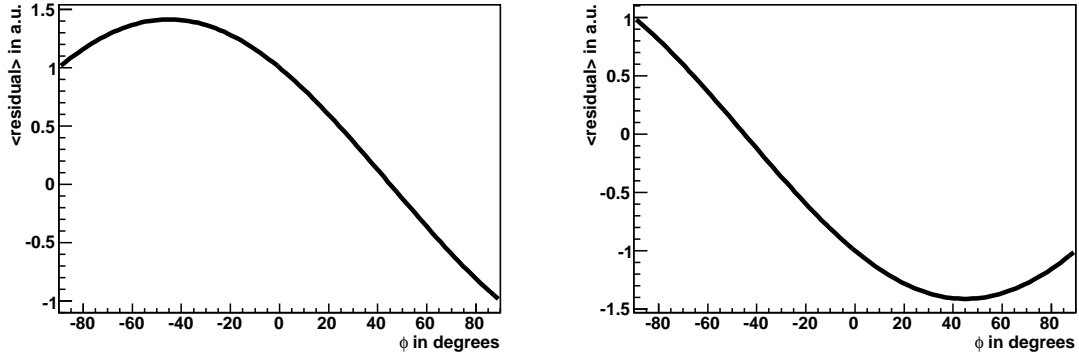}
\caption{Shape of the residual distribution as function of $\phi$ for the sensors as shown in figure~\ref{fig:residuals}.}
\label{fig:shape}
\end{figure}

In order to perform this fit the residuals of both $R$ and $\Phi$ sensors have to be plotted as a function of $\phi$ (and for the $\Phi$ sensor only also as a function of $r$). For the $R$ sensor, the $\phi$ co-ordinate of the residual is taken from the fitted track. Similarly, for the $\Phi$ sensor, the $r$ co-ordinate is taken from the fitted track.

As an example of the resulting distributions, the misalignment introduced in figure~\ref{fig:residuals} would give rise to the shape of the residual distributions shown in figure~\ref{fig:shape}. To keep the fit code simple and general, the fit is performed in the respective VELO half frame. One consequence is that the range in $\phi$ is $[-90^{\circ},90^{\circ}]$ for sensors in the VELO A-side (as shown in figure~\ref{fig:shape}), and $[-180^{\circ},-90^{\circ}]$ and $[90^{\circ},180^{\circ}]$ for sensors in the VELO C-side.

\begin{floatingfigure}[r]{0.45\textwidth}
\centering
\includegraphics[width=0.3\textwidth]{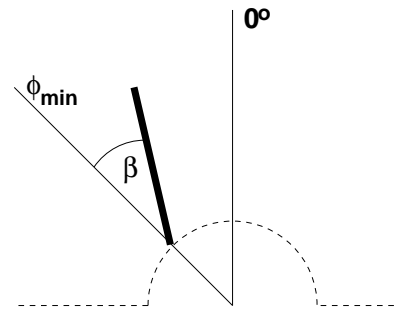}
\caption{Definition of stereo angle.}
\label{fig:stereo}
\end{floatingfigure}

There is a further complication for the VELO $\Phi$ sensors arising from their stereo angle that has been neglected in the discussion above. The strips are not aligned with radial lines but are twisted by a stereo angle. It can be shown that equation~\ref{eqn:residual} for the  $\Phi$ sensor residuals holds if $\phi$ is replaced by $\phi'=\phi_{min}-\beta$, where $\phi_{min}$ is the $\phi$ coordinate at minimum radius of the strip and $\beta$ is the stereo angle, which is defined as the angle between a radial line intersecting the strip at its minimum radius and the strip itself (see fig.~\ref{fig:stereo}).
This relationship holds for both inner and outer region $\phi$ strips, although the sign and magnitude of the stereo angle changes, thus allowing to fit all $\Phi$ sensor hits simultaneously.

Finally, after $\Delta_{x}$,$\Delta_{y}$, and $\Delta_{\gamma}$ have been determined in each iteration, the new alignment constants are applied to the $\Phi$ sensor. For $x$ and $y$ translations, the difference in the misalignment between the $\Phi$ and $R$ sensor is taken as the $\Phi$ sensor's alignment constants, as the common misalignment will be attributed to the module by the module alignment algorithm.
For rotations around the $z$-axis, the alignment constant is taken directly from the $\Phi$ sensor as the $R$ sensor is insensitive to these.

To improve the fit stability a cleaning procedure is applied to the residual distributions.
First, a minimum number of entries is required for the whole distribution (set via job-option, default is 200). 
In order to suppress outliers in the distribution of the residual means, a minimum number of entries is also required for each bin in $\phi$ ($10^{\circ}$).
This threshold is $1/10$ of the threshold for the whole distribution.

To test the validity of the one-dimensional binned fits, a two-dimensional unbinned likelihood fit has been implemented.
No difference has been observed and hence the faster and less complicated one-dimensional fits are kept.

\subsection{Implementation}
In the context of the LHCb alignment software, the VELO sensor alignment algorithm is implemented as part of the {\tt VeloAlignment} package inside the {\tt Alignment} project.
Its iterations are controlled by a python script (located in the job directory of the {\tt Alignment/Escher} package) while all the actual code is a {\tt C++} implementation of a {\tt GaudiTupleAlg}. 

The sensor alignment will be run as the first step of the VELO alignment.
It produces alignment constants that reflect the relative $x$ and $y$ position of the $\Phi$ sensor with respect to the $R$ sensor.
These are then used as input values for the module alignment algorithm.
By definition the $R$ sensor is kept perfectly aligned with the module and hence its alignment constants are all set to $0$.
The output of the module alignment algorithm is then used to update the module alignment constants before aligning the two VELO halves.

As explained below the most time consuming part of this alignment algorithm is the track fit.
Depending on the complexity of the events one iteration of the fit using $20000$ tracks takes about one minute on a single CPU\footnote{1 CPU = 1000 SpecInt2000 units}.

\subsection{Track fits}
For each iteration, unbiased residuals have to be determined from track fits excluding hits on the sensors of the module under study.
This means that the set of hits used for fitting one track will vary when the residuals for sensors of different modules are calculated.
It turns out that the resulting large number of track fits accounts for the bulk of the time consumption of the algorithm.

Two different track fits have been studied.
The bi-directional Kalman track fit as it is used by the main reconstruction software and a straight line track fit, which fits a straight line to a set of at least four space-points made of an $(r/\phi)$ pair.
Both fits show no significant difference in their results for high momentum tracks\footnote{The track sample studied had a flat momentum distribution between $1$ and $100$ GeV.}.
However, the Kalman track fit appears to be roughly $100$ times slower than the straight-line track fit.
A tool is available to extract unbiased residuals from a single Kalman fit using all hits.
As this tool will in the best case leave the Kalman fit a factor $5$ slower than the straight line fit, the latter was used to carry out the larger scale studies presented below.

\section{Simulation studies}
\label{sec:MC}

The sensor alignment method has been tested with 10 samples of randomly generated misalignments with misalignment scales (Gaussian sigmas) of $10~\mathrm{\mu{}m}$ for $x$ and $y$ translations and $1$ mrad for $z$ rotations.
Each sample consists of 20,000 tracks with small slopes ($<1~\mathrm{mrad}$), thus passing through all sensors of one VELO-half and evenly distributed across the sensor surface\footnote{These were generated using the particle gun}. 
Typically three to five iterations of the alignment procedure are required to obtain the best resolution.

To avoid fluctuations in the alignment constants, which were observed when running over minimum bias events, a cut-off has been introduced to avoid updating of the alignment constants if the observed difference to the previous constants is below a certain threshold.
Therefore, the significance of the difference to the previous constants for either $x$ or $y$ misalignments, i.e. the measured difference divided by its estimated error, has to be greater than $1$ (threshold set via job option {\tt MinDeltaSig}).
Its implementation does not alter the performance of the algorithm in the studies as presented below.

\begin{figure}
\centering
    \includegraphics[width=0.47\textwidth]{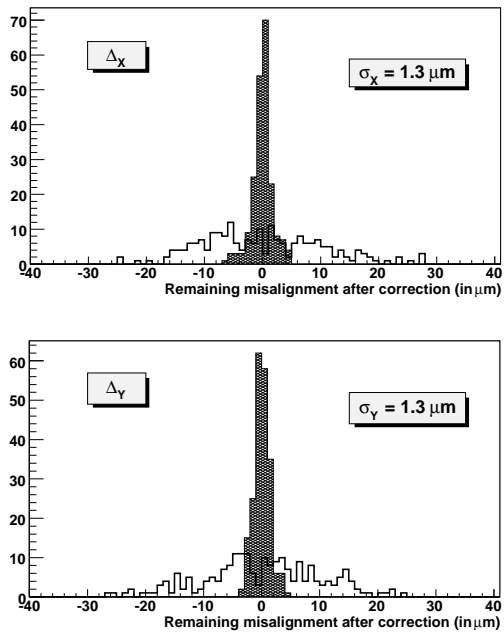}
   \caption{$x$ and $y$ misalignment values before ($\square$), and after ($\blacksquare$) sensor relative alignment.}
   \label{fig:STEP0_result}
\end{figure}

Figure~\ref{fig:STEP0_result} shows the generated and the remaining misalignments after all iterations. The resolution on the relative $x$ and $y$ translation of the sensors of one module is $1.3~\mathrm{\mu{}m}$. This is a significant improvement over the precision expected from the VELO module survey. The performance of this algorithm has also been demonstrated with beam test data and is reported in Ref.~{\bf \cite{bib:NIM-07}}.


It has been shown that the method does not produce a bias when run on a perfect geometry.
However, in case of existing misalignment a bias is observed in some cases.
This is due to the fact that a systematic $z$-dependence of the relative sensor alignment or a common shift will be picked up by all track fits and hence it will be reproduced in the result.
These biases are effectively identical to VELO detector-half tilts or translations and can thus be corrected by the appropriate algorithm.
In practice however, the effective misalignments will most likely be below the sensitivity of this algorithm, and therefore also be negligible in terms of the detector performance.
The results shown in figure~\ref{fig:STEP0_result} have been corrected for effective shifts of the whole detector half, however not for tilts.
In the misalignment samples used here the effective tilts were of the order of a few $\mathrm{\mu{}rad}$.

\begin{figure}[t]
\centering
  \vspace{-2mm}
    \includegraphics[width=0.47\textwidth]{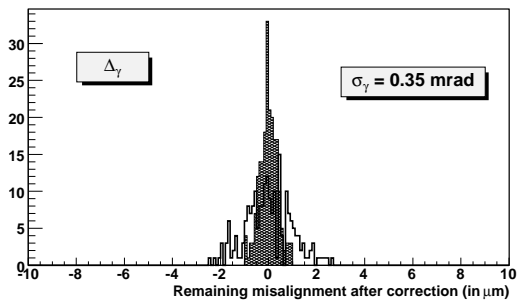}
   \caption{$\gamma$ misalignment values before ($\square$), and after ($\blacksquare$) sensor relative alignment.}
   \label{fig:STEP0_result2}
\end{figure}

Although it will be eventually determined by the module alignment algorithm to a precision of $0.12~\mathrm{mrad}$ \cite{bib:Vir-07}, the extraction of $z$-rotation misalignments has been studied as well.
It both benefits the convergence of the algorithm and provides the module alignment algorithm with a smaller starting problem.
Figure~\ref{fig:STEP0_result2} shows the $z$-rotation alignment constants before and after correction by the algorithm.
The precision turned out to be $0.35~\mathrm{mrad}$.

\section{Conclusion}
\label{sec:concl}

This note describes a method to measure the relative $x$ and $y$ translation misalignments between the $R$ and $\Phi$ sensor on each module from data.
The algorithm has been tested in simulation studies and integrated into the LHCb {\tt Alignment} software project.

The precision achieved for the relative sensor positions is $1.3~\mathrm{\mu{}m}$.
Combined with the precision of the module alignment algorithm, the absolute position of every sensor within either VELO half will be determined to an accuracy of better than $2~\mathrm{\mu{}m}$.

This leaves the sensor position uncertainty due to misalignments well under the best single hit resolution of the sensors.
Hence, the impact of misalignments on LHCb physics should be minimal.

A relative degradation of the resolution of quantities that rely mostly on the VELO performance, e.g. proper time and impact parameter, can be expected to be no more than a few per cent~\cite{bib:Ger-08}.





\end{document}